\PassOptionsToPackage{unicode}{hyperref}
\PassOptionsToPackage{hyphens}{url}
\PassOptionsToPackage{dvipsnames,svgnames,x11names}{xcolor}
\documentclass[
  12pt]{article}

\usepackage{amsmath,amssymb}
\usepackage{iftex}
\ifPDFTeX
  \usepackage[T1]{fontenc}
  \usepackage[utf8]{inputenc}
  \usepackage{textcomp} 
\else 
  \usepackage{unicode-math}
  \defaultfontfeatures{Scale=MatchLowercase}
  \defaultfontfeatures[\rmfamily]{Ligatures=TeX,Scale=1}
\fi
\usepackage{lmodern}
\ifPDFTeX\else  
\fi
\IfFileExists{upquote.sty}{\usepackage{upquote}}{}
\IfFileExists{microtype.sty}{
  \usepackage[]{microtype}
  \UseMicrotypeSet[protrusion]{basicmath} 
}{}
\makeatletter
\@ifundefined{KOMAClassName}{
  \IfFileExists{parskip.sty}{%
    \usepackage{parskip}
  }{
    \setlength{\parindent}{0pt}
    \setlength{\parskip}{6pt plus 2pt minus 1pt}}
}{
  \KOMAoptions{parskip=half}}
\makeatother
\usepackage{xcolor}
\makeatletter
\ifx\paragraph\undefined\else
  \let\oldparagraph\paragraph
  \renewcommand{\paragraph}{
    \@ifstar
      \xxxParagraphStar
      \xxxParagraphNoStar
  }
  \newcommand{\xxxParagraphStar}[1]{\oldparagraph*{#1}\mbox{}}
  \newcommand{\xxxParagraphNoStar}[1]{\oldparagraph{#1}\mbox{}}
\fi
\ifx\subparagraph\undefined\else
  \let\oldsubparagraph\subparagraph
  \renewcommand{\subparagraph}{
    \@ifstar
      \xxxSubParagraphStar
      \xxxSubParagraphNoStar
  }
  \newcommand{\xxxSubParagraphStar}[1]{\oldsubparagraph*{#1}\mbox{}}
  \newcommand{\xxxSubParagraphNoStar}[1]{\oldsubparagraph{#1}\mbox{}}
\fi
\makeatother

\usepackage{longtable,booktabs,array}
\usepackage{calc} 
\usepackage{etoolbox}
\makeatletter
\patchcmd\longtable{\par}{\if@noskipsec\mbox{}\fi\par}{}{}
\makeatother
\IfFileExists{footnotehyper.sty}{\usepackage{footnotehyper}}{\usepackage{footnote}}
\makesavenoteenv{longtable}
\usepackage{graphicx}
\makeatletter
\def\maxwidth{\ifdim\Gin@nat@width>\linewidth\linewidth\else\Gin@nat@width\fi}
\def\maxheight{\ifdim\Gin@nat@height>\textheight\textheight\else\Gin@nat@height\fi}
\makeatother
\setkeys{Gin}{width=\maxwidth,height=\maxheight,keepaspectratio}
\makeatletter
\def\fps@figure{htbp}
\makeatother

\makeatletter
\@ifpackageloaded{caption}{}{\usepackage{caption}}
\AtBeginDocument{%
\ifdefined\contentsname
  \renewcommand*\contentsname{Table of contents}
\else
  \newcommand\contentsname{Table of contents}
\fi
\ifdefined\listfigurename
  \renewcommand*\listfigurename{List of Figures}
\else
  \newcommand\listfigurename{List of Figures}
\fi
\ifdefined\listtablename
  \renewcommand*\listtablename{List of Tables}
\else
  \newcommand\listtablename{List of Tables}
\fi
\ifdefined\figurename
  \renewcommand*\figurename{Figure}
\else
  \newcommand\figurename{Figure}
\fi
\ifdefined\tablename
  \renewcommand*\tablename{Table}
\else
  \newcommand\tablename{Table}
\fi
}
\@ifpackageloaded{float}{}{\usepackage{float}}
\floatstyle{ruled}
\@ifundefined{c@chapter}{\newfloat{codelisting}{h}{lop}}{\newfloat{codelisting}{h}{lop}[chapter]}
\floatname{codelisting}{Listing}

\makeatother
\makeatletter
\@ifpackageloaded{caption}{}{\usepackage{caption}}
\@ifpackageloaded{subcaption}{}{\usepackage{subcaption}}
\makeatother

\ifLuaTeX
  \usepackage{selnolig}  
\fi
\usepackage[]{natbib}
\usepackage{bookmark}

\IfFileExists{xurl.sty}{\usepackage{xurl}}{} 
\hypersetup{
  pdftitle={Title},
  pdfauthor={Author 1; Author 2},
  pdfkeywords={3 to 6 keywords, that do not appear in the title},
  colorlinks=true,
  linkcolor={blue},
  filecolor={Maroon},
  citecolor={Blue},
  urlcolor={Blue},
  pdfcreator={LaTeX via pandoc}}

\newcommand{\anon}{1}

\usepackage{enumitem}
\newtheorem{theorem}{Theorem} 
\newtheorem{definition}[theorem]{Definition}
\usepackage{cancel}
\usepackage{float} 
\usepackage{algorithm}
\usepackage{algorithmic}
\usepackage{amsmath}     
\usepackage{amssymb}    
\usepackage{bm}
\usepackage{multirow}
\begin{document}

\def\spacingset#1{\renewcommand{\baselinestretch}%
{#1}\small\normalsize} \spacingset{1}


\if1\anon
{
  \title{\bf Optimizing Real-Time Oxytocin Administration to Prevent Postpartum Hemorrhage: A Bayesian Approach to Dynamic Treatment Regimes}
  \author{Haiyan Zhu, Yuqi Qiu$^{*}$, and Yingchun Zhou$^{*}$ \\
	Key Laboratory of Advanced Theory and Application in Statistics \\
	and Data Science - MOE, School of Statistics,\\ 
	East China Normal University, 200062, Shanghai, China.\\
	{\itshape{Email}}: $^*$yqqiu@fem.ecnu.edu.cn\\
                       $^*$yczhou@stat.ecnu.edu.cn}
  \maketitle
} \fi

\if0\anon
{
  \bigskip
  \bigskip
  \bigskip
  \begin{center}
    {\LARGE\bf Optimizing Real-Time Oxytocin Administration to Prevent Postpartum Hemorrhage: A Bayesian Approach to Dynamic Treatment Regimes}
\end{center}
  \medskip
} \fi

\bigskip
\begin{abstract}
Postpartum hemorrhage (PPH) remains a leading cause of maternal morbidity and mortality worldwide. Oxytocin, though widely recognized for facilitating labor, is also the primary pharmacological intervention for PPH prevention. However, current dosing protocols lack personalization and fail to account for real-time physiological changes during labor. Moreover, standard dynamic treatment regime (DTR) methods cannot accommodate the continuous monitoring and adjustment. To address this, we propose a semiparametric Bayesian method for estimating an optimal treatment regime in real-time, which allows for the existence of latent individual-level variables. Specifically, random real-time DTRs are defined through interventional parameters, optimized by minimizing posterior predictive loss. We further introduce a "physician-in-the-loop" framework to align optimal strategies with clinical expertise. In an application to Consortium on Safe Labor data, the proposed method achieved consistently lower estimated blood loss than other competing methods. The learned policy recommends earlier initiation, rapid dose escalation, and more frequent titration for parturients with higher BMI, alongside increased adjustments relative to cervical dilation and the interval since the last dose change. Simulation studies demonstrate robust performance and computational efficiency, especially when unmeasured patient factors influence outcomes and covariates. Supplementary materials provides a standardized description of the materials available for reproducing the work.
\end{abstract}

\noindent%
\vfill

\newpage
\spacingset{1.8} 

\section{Introduction}
\label{sec:intro}
\begin{figure}
	\centering
	\includegraphics[width=0.65\linewidth]{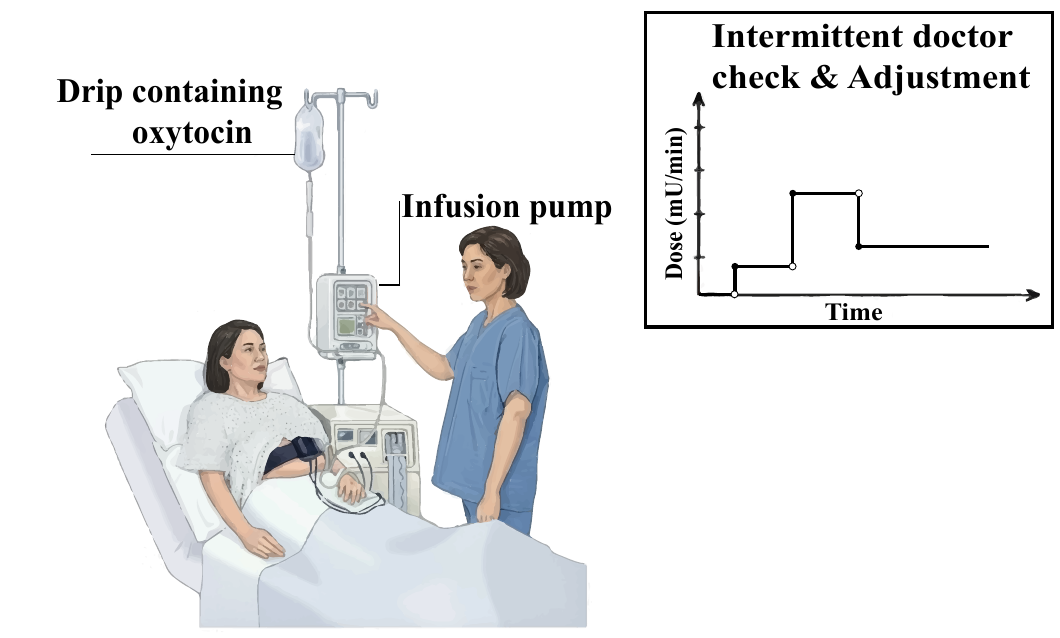}
	\caption{Current oxytocin administration scheme.}
	\label{fig:Intro_back}
\end{figure}

Postpartum hemorrhage (PPH), commonly defined as blood loss exceeding 500 mL following vaginal delivery or 1,000 mL after cesarean section, accounts for approximately one-quarter of maternal deaths worldwide \citep{tunccalp2013new, Wang2025,say2014}. 
Oxytocin, a potent uterotonic agent, serves as the first-line pharmacological intervention for PPH prevention \citep{roach2013dose,lambert2020oxytocin}, a standard endorsed by the World Health Organization \citep{world2012recommendations}. 
However, determining the optimal timing, dose, and titration strategy for oxytocin administration remains challenging. Current protocols rely largely on population-level guidelines, leading clinicians to adjust dose levels intermittently without fully accounting for individual heterogeneity or continuously changing maternal status (Figure \ref{fig:Intro_back}). For instance, a parturient with elevated body mass index (BMI) at 38 weeks gestation undergoes labor augmentation with oxytocin. Should the dose be increased earlier compared to a parturient with normal BMI? How should cervical dilation patterns inform real-time adjustments? Such clinical scenarios demand personalized, data-driven decision rules that adapt dynamically to each parturient's evolving condition.

Although observational studies indicate that factors such as BMI and cervical dilation predict oxytocin requirements \citep{carlson2017oxytocin}, existing evidence remains qualitative and lacks actionable guidance on the magnitude and frequency of real-time dosing. Clinical decision support systems (CDSS) offer a mechanism to translate such evidence into practice, similar to their deployment in diabetes and cancer pain management \citep{castle2017future, smith2004pain}. However, effective CDSS development requires methodological advances capable of handling continuous monitoring and unmeasured patient heterogeneity. Moreover, such systems must operate as "physician-in-the-loop" frameworks, supporting rather than replacing clinical expertise \citep{shortliffe2018clinical}. Therefore, developing a theoretical framework for optimal real-time decision recommendation that integrates professional judgment is of significant theoretical and practical importance.

\subsection{Dynamic Treatment Regimes: From Discrete to Continuous Time}
\label{subsec:dtr_background}

Dynamic treatment regimes (DTRs), also known as adaptive treatment strategies or treatment policies, formalize sequential decision rules that personalize treatment based on patients' time-varying states \citep{murphy2003optimal,chakraborty2013statistical,chakraborty2014dynamic}. In discrete-time settings with fixed decision points, substantial progress has been made using frequentist approaches, such as outcome-weighted learning \citep{zhao2012estimating}, Q-learning \citep{zhao2009reinforcement}, A-learning \citep{murphy2003optimal}, and doubly robust estimators \citep{zhang2013robust}. On the Bayesian front, \cite{arjas2010optimal} combined nonparametric regression with backward induction; \cite{saarela2015predictive} developed predictive inference frameworks; \cite{murray2018bayesian} adapted Q-learning via Bayesian machine learning; and \cite{rodriguez2023semiparametric} proposed semiparametric methods using dynamic marginal structural models. Nevertheless, these methods assume a small, fixed number of decision points, which is an assumption incompatible with real-time clinical scenarios where treatment adjustments occur continuously in response to ongoing physiological monitoring.

Extending DTR frameworks to continuous time presents challenges, as treatment histories become functional processes requiring optimization over infinite-dimensional decision spaces. \cite{Hua_2022} represented one of the few attempts to address continuous-time DTRs via a parametric Bayesian joint model. However, their focus on scheduled medical visits differs substantively from the instantaneous treatment decisions required in real-time support systems. Moreover, parametric approaches necessitate correct specification of all intermediate variable models, increasing both computational burden and the risk of model misspecification in the presence of unmeasured covariates.

\subsection{Contributions}
\label{subsec:contributions}

This paper studies real-time oxytocin titration during labor, where dosing is adjusted repeatedly in response to evolving clinical status. Thus, we propose a semiparametric Bayesian framework for learning optimal random real-time DTRs in the presence of latent individual-level heterogeneity. 

Our primary contributions include: (1) Development of a "physician-in-the-loop" optimal real-time dose recommendation system designed to incorporate clinical expertise and treatment preferences (Figure \ref{fig:Optimal_Real_Time_Recom_System}). (2) Applying our method to a real-world oxytocin administration dataset from the Consortium on Safe Labor (CSL), we provide real-time optimal oxytocin dose recommendations and demonstrate a policy-relevant population health impact. Overall, the optimal regimen shows that higher maternal BMI necessitates earlier initiation, higher dosages, and more frequent monitoring. The optimized regimens reduce the mean estimated blood loss from 321 mL under current practice to 115–150 mL across oxytocin thresholds, corresponding to a 53–64\% relative reduction. If similar reductions in blood loss were achieved among high-risk parturients at scale, they could contribute meaningfully to ongoing efforts to reduce maternal morbidity and mortality from hemorrhage. (3) On the methodological side, we formally define real-time random DTRs through interventional parameters $\bm{\eta}$ that govern treatment switching intensities $\lambda^A_{\mathcal{E}}(t \mid \mathcal{H}^A_t, \bm{\eta})$, and develop a semiparametric inference framework to estimate the optimal $\bm{\eta}$. Unlike existing parametric methods, our framework permits unmeasured individual-level variables $\bm{U}$ to influence both outcomes and time-varying covariates, provided $\bm{U}$ remains conditionally independent of treatment assignments given observed histories. This relaxation proves critical in obstetric applications, where latent physiological drivers (e.g., intrinsic uterine contractility) are commonly encountered. (4) On the computational front, our method achieves efficiency through targeted optimization. Unlike existing approaches that require estimating the entire nuisance parameters, we focus on optimizing the interventional parameters $\bm{\eta}$, significantly reducing computational burden.


\subsection{Organization}
\label{subsec:organization}

The remainder of this article proceeds as follows.
Section~\ref{Physician-in-the-Loop_Optimal_Dose_Guidance} introduces the proposed optimal real-time dose recommendation system, providing a "physician-in-the-loop" solution for optimal decision-making. Section~\ref{Methodology} details the estimation procedure of optimal real-time dose-stratum switching strategy, which is the core component of the recommendation system: Section~\ref{Notation} introduces notation and Section~\ref{Modeling_Framework} describes the two-world paradigm (observed vs.\ experimental), distinguishing between the data-generating mechanism under current practice and the counterfactual mechanism under optimized policies; Section~\ref{Random_Real-Time_DTRs} formalizes random real-time DTRs through interventional parameters and establishes identifiability; Section~\ref{Optimal_Real-Time_DTRs} derives the Bayesian decision rule for obtaining optimal parameters by minimizing posterior predictive loss. Section~\ref{Application} applies our framework to oxytocin administration data from the CSL. Section~\ref{Simulation} validates the proposed method through simulation studies under four scenarios, emphasizing performance gains when latent confounders are present. This is precisely the setting where parametric competitors struggle due to model misspecification.  Section~\ref{Discussion} reflects on main findings, discusses limitations regarding model specifications, and outlines extensions to multi-category treatments. Supplementary materials provide theoretical proofs, data generation algorithm, details of real dataset analysis, and annotated code for reproducibility.

\section{Physician-in-the-Loop Optimal Dose Guidance}
\label{Physician-in-the-Loop_Optimal_Dose_Guidance}

\begin{figure}[hbpt]
	\centering
	\includegraphics[width=0.89\linewidth]{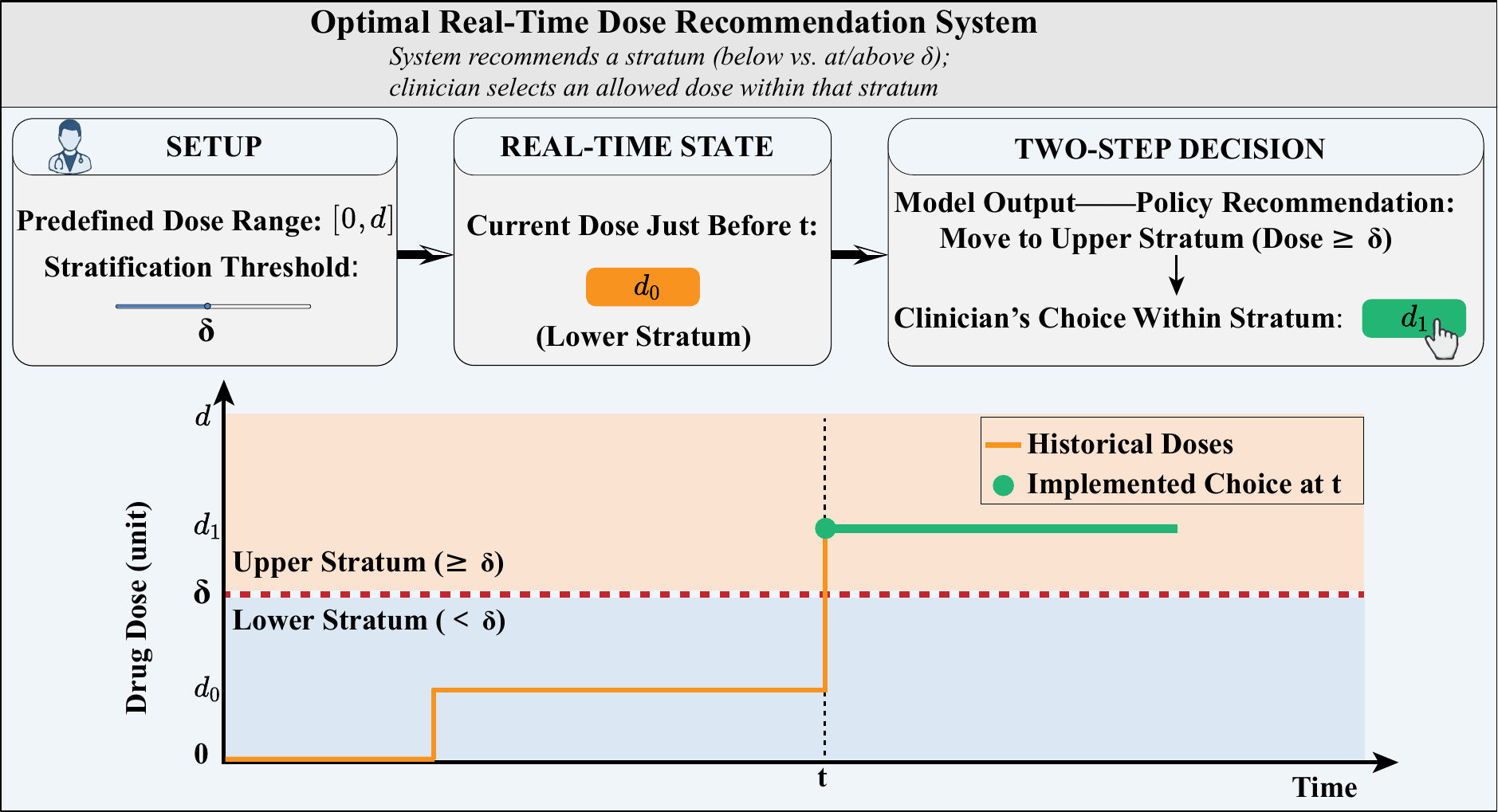}
	\caption{Schematic of the proposed optimal real-time dose recommendation system. In the case where $\delta=0$, the upper stratum corresponds to positive doses ($>0$), while the lower stratum represents zero dose ($=0$).}
	\label{fig:Optimal_Real_Time_Recom_System}
\end{figure}

Traditional "human-in-the-loop" paradigm in AI systems typically position humans as data-labeling oracles \citep{settles2009active} or sources of domain knowledge \citep{mosqueira2023human}. In clinical decision settings, however, physicians function as ultimate decision-makers, possessing expertise and patient-specific insights that are difficult to fully encode algorithmically. Consequently, we propose a "physician-in-the-loop" framework that integrates algorithmic optimization with clinical authority (Figure~\ref{fig:Optimal_Real_Time_Recom_System}). 

Instead of prescribing rigid doses that may conflict with clinical judgment, our system operates through a collaborative hierarchical decision-making process: (1) Physicians first specify a clinically acceptable dose range and select an appropriate dose stratification threshold based on patient assessment; (2) The system then recommends an optimal dose stratum to use (below the threshold versus at/above the threshold) in real-time; (3) The final specific dose is selected by the physician from the pre-specified range within the algorithm recommended stratum.

The primary objective of this system is to reduce the burden of continuous bedside monitoring and frequent manual dose changes across many patients. In practice, the physician uses clinical judgment to specify a dose range for a short upcoming time window. During this window, the proposed algorithm recommends a dose stratum in real-time, from which the physician selects a specific dose. The length of the window can be chosen to match clinical workflow. As the window ends, the physician updates the dose range for the next window so that dosing remains continuous.

Regarding the selection of the stratification threshold $\delta$, priority should be given to clinical domain knowledge when available. Alternatively, in the absence of such prior information, the median of the patient-specific dose range represents a sensible option for $\delta$. Notably, our empirical analysis demonstrates that the proposed method consistently optimizes outcomes using several stratification thresholds across a range.

To implement this "physician-in-the-loop" optimal dose recommendation system, the recommendation strategy can be integrated into standard infusion devices, which is technically feasible. Algorithmically, the key problem is to estimate an optimal real-time treatment switching policy from data; this is the focus of Section~\ref{Methodology}.

\section{Methodology}
\label{Methodology}

In this section, we introduce how to estimate the optimal real-time dose-stratum switching strategy. This strategy is the key algorithmic component of the "physician-in-the-loop" dose recommendation system proposed in Section~\ref{Physician-in-the-Loop_Optimal_Dose_Guidance}.

\subsection{Notation}
\label{Notation}
Consider a longitudinal observational study with $n$ independent units. For each unit, we observe $(T_{\max}, \bar{A}, \bar{\bm{L}}, Y)$ and allow for latent variables $\bm{U}$, where 
\begin{itemize}[leftmargin=15pt]
	\item $T_{max}\in (0,t_R]$  denotes the end time of treatment within a finite follow-up window.
	\item $\bar{A}=\{A(t):t\in [0,T_{max}]\}$ denotes the whole treatment process, where $A(t)\in\{0,1\}$ denotes the binary treatment assignment at time $t$ ($1$ for upper stratum, $0$ for lower stratum). The definition of the "upper stratum"  can be flexibly chosen by the clinician (e.g., $\geq 4$ mU/min for oxytocin). Let $\Delta_A(t) = 1\{A(t) \neq A(t^-)\}$ indicate a change in stratum at time $t$.
	\item $\bar{\bm{L}}=\{\bm{L}(t):0\le t\le T_{max}\}$ denotes the entire covariate process, with $\bm{L}(t)\in\mathbb{R}^{p_L}$ containing all baseline and time-varying confounders. To eliminate unnecessary technical distractions, time-varying confounders in $\bm{L}(t)$ are assumed to change only at monitoring times. This assumption imposes no serious practical limitations, and the flexibility of $\bm{L}(t)$ will increase correspondingly with the sample size. For individual $i$, if $\bm{L}_i(t)$ is observed at the time points $\{0 = t_0 < t_1 < \cdots < t_m < t_{m+1} = T_{max,i}\}$, then for any $t \in [t_k, t_{k+1})$ where $k \in \{0, \dots, m\}$, $\bm{L}_i(t)$  is defined as $\bm{L}_i(t_k)$.
	\item $Y\in\mathbb{R}$ is the outcome measured at the end of study. 
	\item  $\bm{U}\in\mathbb{R}^{p_U}$ denotes latent factors that may influence both outcome and intermediate variables, but is assumed conditionally independent of treatment assignments.
\end{itemize}
Moreover, we use the overbar notation to represent the history of a random variable (e.g., $\bar{A}(t)=\{A(s):0\le s \le t\}$ denotes the treatment history up to and including time $t$), and let $\mathcal{H}^{A}_t=\{\bar{A}(t^-),\bar{\bm{L}}(t),t\le T_{max}\}, \mathcal{H}^{L}_t=\{\bar{A}(t^-),\bar{\bm{L}}(t^-),\bm{U},t\le T_{max}\}, \mathcal{H}^{T_{max}}_t=\{\bar{A}(t^-),\bar{\bm{L}}(t^-),\bm{U},t^-<T_{max}\}$ be the history information that influence $A(t),\bm{L}(t),N^{T_{max}}(t)$, respectively. 
The notation $t^-$ denotes the left-hand limit at time $t$. $N^{T_{max}}(t)$, the zero-one counting process (\citealt{andersen2012statistical}), indicates treatment completion.

\subsection{Causal Framework and the Two-World Paradigm}
\label{Modeling_Framework}
To learn an optimal real-time switching policy from observational data, we require three causal assumptions that generalize standard discrete-time conditions \citep{robins1999association} to continuous-time settings:\\
{\it{Continuous-Time Consistency.}} Let $Y_{\bar{a}}$ denote the potential outcome under a treatment regime $\bar{a}$, and assume that for any treatment regime $\bar{a}$, $Y=Y_{\bar{a}}$, almost surely. Mirroring discrete-time scenarios, the consistency assumption connects observed outcomes to potential outcomes via the actually received treatment process. Specifically, when an individual receives treatment $\bar{a}$, his/her observed outcome $Y$ equals $Y_{\bar{a}}$. 

\textit{Continuous-time sequential ignorability.} Let $N^A(t)$ be the counting process that records the number of treatment stratum changes up to time $t$, and let $\lambda^A(t)$ denote its (predictable) intensity. For example, $N^A(t)=2$ means the treatment stratum has changed twice by time $t$. We assume that, conditional on the observed history, $\lambda^A(t\mid\mathcal{H}^A_t,Y_{\bar{a}})=\lambda^A(t\mid\mathcal{H}^A_t)$ so the instantaneous switching rate at time $t$ does not depend on the counterfactual outcome once the observed treatment and covariate history is given.

\textit{Positivity.} We assume there is sufficient variation in observed treatment switching to support the candidate policies considered. In particular, for histories that occur with positive probability, the switching intensity under the observed practice is positive whenever the candidate policy assigns positive probability to a switch. We discuss the practical implications and diagnostics in Section~\ref{Optimal_Real-Time_DTRs}.

\begin{figure}[htbp]
	\centering
	\includegraphics[width=0.83\linewidth]{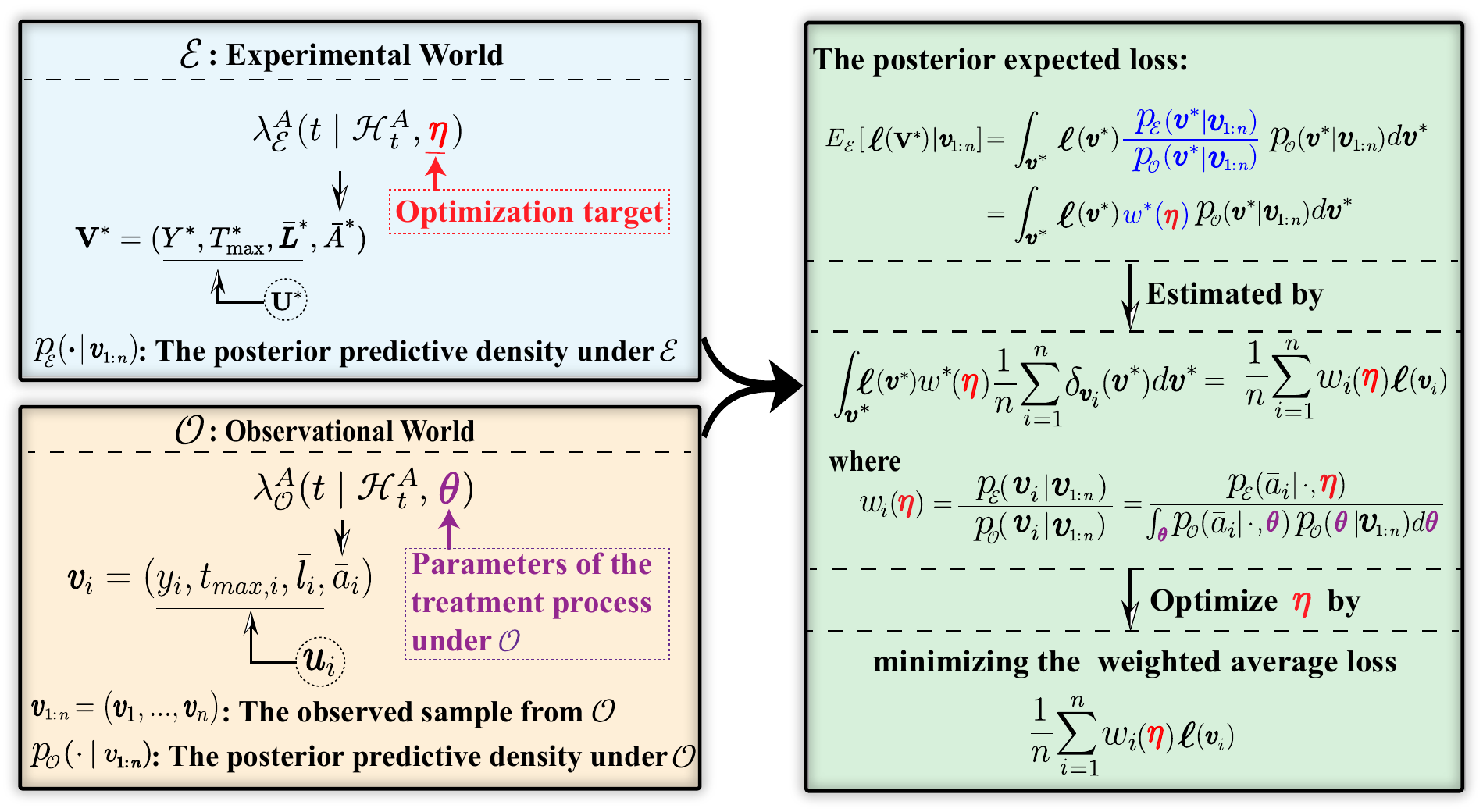}
	\caption{Workflow of the proposed method for optimal random real-time DTRs. Data originates from the observational world $\mathcal{O}$ (current practice), while the goal is to identify optimal interventional parameters $\bm{\eta}$ defining policies in the experimental world $\mathcal{E}$. Both worlds share outcome, covariate, and latent variable models; only treatment intensity differs.}
	\label{fig:Technical Roadmap}
\end{figure}

\textbf{The Two-World Framework.} Figure~\ref{fig:Technical Roadmap} depicts our approach. We consider two settings. In the \textit{observational world} $\mathcal{O}$, treatment switching follows current clinical practice, modeled through an intensity $\lambda^A_{\mathcal{O}}(t \mid \mathcal{H}^A_t,\bm{\theta})$ with parameters $\bm{\theta}$. In the \textit{experimental world} $\mathcal{E}$, where intensities $\lambda^A_{\mathcal{E}}(t \mid \mathcal{H}^A_t, \bm{\eta})$ are controlled by interventional parameters $\bm{\eta}$ that we aim to optimize. 

The two worlds share identical data-generating mechanisms for outcomes, covariates, and latent variables. They differ only in the treatment allocation rule. This structure enables us to use observed data from $\mathcal{O}$ to make inferences about optimal policies in $\mathcal{E}$ via importance weighting. Intuitively, we reweight observed losses according to how likely each patient's treatment history would be under the experimental policy $\bm{\eta}$ versus the observational policy $\bm{\theta}$, then select $\bm{\eta}$ to minimize the weighted average loss.

With these notations and causal setup in place, we next define randomized real-time DTRs in Section \ref{Random_Real-Time_DTRs} and then describe the procedures for estimating the optimal randomized real-time DTRs in Section \ref{Optimal_Real-Time_DTRs}.

\subsection{Random Real-Time DTRs}
\label{Random_Real-Time_DTRs}
To define DTRs in continuous time, we must specify how treatment can change over an interval rather than at a fixed set of decision points. In a discrete-time setting with $K$ decision times, a deterministic regime is a collection of rules $d_k:\mathcal{H}_k\to\{0,1\}$ that map the observed history to a treatment choice. However, in continuous time with uncountably infinite decision points $t \in [0, t_R]$, specifying deterministic rules for each $t$ is not practical. In addition, fully deterministic continuous-time rules tend to create weak overlap thus violate positivity: it becomes unlikely to observe treatment paths that match any single deterministic regime closely, which makes estimation unstable \citep{rodriguez2023semiparametric}.

For these reasons, we focus on \textit{random} real-time DTRs, which specify treatment switching through an intensity function. Randomization also has practical value: it avoids a hard decision boundary and allows controlled exploration of treatment options when multiple choices are clinically acceptable \citep{murphy2001marginal,sutton1998reinforcement}.

\begin{definition}[Random Real-Time DTRs]\label{def1}
	Let $\bm{\eta}$ be an interventional-parameter vector.  A random real-time DTR in the experimental world $\mathcal{E}$ is indexed by $\bm{\eta}$ and defined through a treatment switching intensity $\lambda^A_{\mathcal{E}}(t\mid \mathcal{H}^A_t,\bm{\eta})$ that satisfies, for all relevant $t$,:
	\[
	0 < \lambda^A_{\mathcal{E}}(t\mid \mathcal{H}^A_t,\bm{\eta}) < \infty,\]
	where $\mathcal{H}^A_t=\{\bar{A}(t^-),\bar{\bm{L}}(t),0\le t\le T_{max}\}$.
\end{definition}

Definition~\ref{def1} extends the usual DTR concept to continuous time by replacing pointwise decision rules with a switching-rate model. The interventional-parameter vector $\bm{\eta}$ determines how the switching intensity depends on the observed history, and the condition above ensures the switching intensity is positive and finite over the treatment period. It is worth noting that the observational practice model \(\lambda^A_{\mathcal{O}}(t\mid \mathcal{H}^A_t,\bm{\theta})\in (0,\infty)\) can be viewed as a special case of Definition~\ref{def1} by taking $\mathcal{E}=\mathcal{O}$.

Under Definition \ref{def1}, the problem of finding the optimal real-time DTR $d$ then reduces to identifying the optimal parameters $\bm{\eta}$. Section \ref{Optimal_Real-Time_DTRs} obtains the optimal $\bm{\eta}$ by minimizing the posterior predictive loss through a Bayes decision rule.

\subsection{Optimal Random Real-Time DTRs}
\label{Optimal_Real-Time_DTRs}
Consider a sequential decision problem with uncountable infinite decision points in a closed interval and a final outcome $Y$ to be observed when treatment ends. As discussed in Section \ref{Random_Real-Time_DTRs}, we work with \emph{random} real-time DTRs defined through a treatment switching intensity. Under Definition \ref{def1}, finding the optimal random real-time DTR $d$ is equivalent to finding the optimal interventional parameters $\bm{\eta}$.

For unit $i$, let $\bm{v}_i=(t_{\max,i},\bar{a}_i,\bar{\bm{l}}_i,y_i)^\top$, $i=1,\ldots,n$, denote the observed data from the observational world $\mathcal{O}$ (latent variables $\bm{u}_i$ are not observed). Our two-world setup assumes that the observational and experimental worlds share the same data-generating mechanisms for $(Y,T_{\max},\bar{\bm{L}},\bm{U})$ and differ only in the treatment switching intensity. This implies that the distribution under $\mathcal{E}$ can be related to that under $\mathcal{O}$ by a likelihood ratio that depends only on the treatment switching model. A formal derivation is given in \textbf{Supplementary Material A-B}.

As in \citet{Saarela_2015}, we assume that $(\bm{v}^\top_i,\bm{u}^\top_i)^\top$ are infinitely exchangeable over the unit indices $i=1,2,...$. Deducing the De Finetti representation \citep{bernardo1994bayesian}, the joint distribution in the observational world $\mathcal{O}$ can be factorized as
\begin{align*}
	p_{\mathcal{O}}(\bm{v}_{1:n})&=\int_{\bm{\phi,\theta},\bm{u}_{1:n}}p_{\mathcal{O}}(\bm{v}_{1:n},\bm{u}_{1:n}\mid\bm{\phi,\theta})p(\bm{\phi,\theta})d\bm{u}_{1:n}d\bm{\phi} d\bm{\theta}\nonumber\\
	&=\int_{\bm{\phi,\theta}}\prod\limits_{i=1}^n\left[\int_{\bm{u}_i}p(y_i\mid \cdot,\bm{u}_i,\bm{\phi}_Y)p(t_{max,i}\mid\cdot,\bm{u}_i,\bm{\phi}_{T_{max}})p(\bar{\bm{l}}_i\mid\cdot,\bm{u}_i,\bm{\phi}_L)p(\bm{u}_i\mid\bm{\phi}_U)d\bm{u}_i\right.\nonumber\\
	&\quad\left.\times p_{\mathcal{O}}(\bar{a}_i\mid \cdot,\bm{\theta})\right]p(\bm{\phi},\bm{\theta})d\bm{\phi} d\bm{\theta},
\end{align*}
where $\bm{u}_{1:n}=(\bm{u}^\top_1,\dots,\bm{u}^\top_n)^\top$, $\bm{\phi}=(\bm{\phi}^\top_Y,\bm{\phi}^\top_{T_{max}},\bm{\phi}^\top_L,\bm{\phi}^\top_U)^\top$ is a partitioning of the non-interventional parameters. $p(y_i\mid \cdot,\bm{u}_i,\bm{\phi}_Y)$, $p(t_{max,i}\mid\cdot,\bm{u}_i,\bm{\phi}_{T_{max}})$ and $p(\bar{\bm{l}}_i\mid\cdot,\bm{u}_i,\bm{\phi}_L)$ represent the likelihood decompositions of $Y$,  $T_{max}$ and $\bar{\bm{L}}$ for the i-th individual, respectively. See \textbf{Supplementary Material A} for more details. The likelihood decomposition of $\bar{a}_i$, $p_{\mathcal{O}}(\bar{a}_i\mid \cdot,\bm{\theta})$, is defined as
$$\prod\limits_{t_j:\Delta_{a_i}(t_j)=1}\lambda^A_{\mathcal{O}}(t_j\mid\mathcal{H}^{A}_{it_j},\bm{\theta})\exp\{-\int_0^{t_{max,i}}\lambda^A_{\mathcal{O}}(t\mid\mathcal{H}^{A}_{it},\bm{\theta})dt\}.$$ Here, the first component is given by a finite product over jump times, while the second component represents the survival function for treatment changing (ref. \citet{Hu_et_al_2023}).

In the experimental world $\mathcal{E}$, assume infinite exchangeability, the outcome, covariate, and latent-variable models remain the same, but the switching intensity is uniquely determined by the fixed interventional parameter vector $\bm{\eta}$:
\begin{equation*}
	\begin{aligned}
		p_{\mathcal{E}}(\bm{v}_{1:n})&=\int_{\bm{\phi},\bm{u}_{1:n}}p_{\mathcal{E}}(\bm{v}_{1:n},\bm{u}_{1:n}\mid\bm{\phi})p(\bm{\phi})d\bm{u}_{1:n}d\bm{\phi}\\
		&=\int_{\bm{\phi}}\prod\limits_{i=1}^n\left[\int_{\bm{u}_i}p(y_i\mid \cdot,\bm{u}_i,\bm{\phi}_Y)p(t_{max,i}\mid\cdot,\bm{u}_i,\bm{\phi}_{T_{max}})p(\bar{\bm{l}}_i\mid\cdot,\bm{u}_i,\bm{\phi}_L)p(\bm{u}_i\mid\bm{\phi}_U)d\bm{u}_i\right]p(\bm{\phi})d\bm{\phi}\\
		&\quad\times p_{\mathcal{E}}(\bar{a}_i\mid \cdot,\bm{\eta}),
	\end{aligned}\label{equationJd}
\end{equation*}
where $p_{\mathcal{E}}(\bar{a}_i\mid \cdot,\bm{\eta})$ is defined as
$$\prod\limits_{t_j:\Delta_{a_i}(t_j)=1}\lambda^A_{\mathcal{E}}(t_j\mid\mathcal{H}^{A}_{it_j},\bm{\eta})\exp\{-\int_0^{t_{max,i}}\lambda^A_{\mathcal{E}}(t\mid\mathcal{H}^{A}_{it},\bm{\eta})dt\}.$$
To find the optimal $\bm{\eta}$, one can minimize the posterior expected loss $E_{\mathcal{E}}[\ell(\bm{V}^*)\mid \bm{v}_{1:n}]$, where $\ell(\cdot)$ is a loss function, $\bm{V}^*=(T_{max}^*,\bar{A}^*,\bar{\bm{L}}^*,Y^*)^\top$ is a random observation sampled from the super-population or data-generating mechanism characterized by $p_{\mathcal{E}}(\bm{v}_{1:n})$. Specifically,
$$
\begin{aligned}
	E_{\mathcal{E}}[\ell(\bm{V}^*)\mid \bm{v}_{1:n}]&=\int_{\bm{v}^*}\ell(\bm{v}^*)p_{\mathcal{E}}(\bm{v}^*\mid \bm{v}_{1:n})d\bm{v}^*\\
	&=\int_{\bm{v}^*}\ell(\bm{v}^*)\frac{p_{\mathcal{E}}(\bm{v}^*\mid \bm{v}_{1:n})}{p_{\mathcal{O}}(\bm{v}^*\mid \bm{v}_{1:n})}p_{\mathcal{O}}(\bm{v}^*\mid \bm{v}_{1:n})d\bm{v}^*\\
	&:=\int_{\bm{v}^*}w^*(\bm{\eta})\ell(\bm{v}^*)p_n(\bm{v}^*)d\bm{v}^*,
\end{aligned}
$$ 
where $\bm{v}^*$ is a realization of $\bm{V}^*$, $w^*(\bm{\eta})=\frac{p_{\mathcal{E}}(\bm{v}^*\mid \bm{v}_{1:n})}{p_{\mathcal{O}}(\bm{v}^*\mid \bm{v}_{1:n})}$, and $p_n(\bm{v}^*)$ is taken to be a non-parametric predictive density. The weight $w^*(\bm{\eta})$ is well-defined ({\it{Positivity}}). Formally, we require absolute continuity of the experimental measure with respect to the observational measure (cf. \citet{Saarela_2015}).

In practice, researchers can choose different loss functions $\ell(\bm{V}^*)$ and predictive densities $p_n(\bm{v}^*)$ based on what scientific question they aim to address. Specifically, we define $\ell(\bm{V}^*) = \exp(Y^*)$ to align with the real dataset. In our real dataset, $Y = \log(\text{EBLoss})$, where EBLoss denotes the estimated blood loss. The log-transformation induces approximate normality, and $\exp(Y)$ recovers the original scale. Additionally,  we set $p_n(\bm{v}^*)=\frac{1}{n}\sum\limits_{k=1}^n\delta_{\bm{v}_k}(\bm{v}^*)$, where $\delta_x(y)=1\{y=x\}$ is an indicator function. Then $E_{\mathcal{E}}[\exp(Y^*)\mid \bm{v}_{1:n}]$ can be approximated by
$$\frac{1}{n}\sum\limits_{i=1}^nw_i(\bm{\eta})\exp(y_i).$$

The weight used in the empirical approximation takes the form:
$$w_i(\bm{\eta})=\frac{\prod\limits_{t_j:\Delta_{a_i}(t_j)=1}\lambda^A_{\mathcal{E}}(t_j\mid\mathcal{H}^{A}_{it_j},\bm{\eta})\exp\{-\int\lambda^A_{\mathcal{E}}(t\mid\mathcal{H}^{A}_{it},\bm{\eta})dt\}}{E_{\bm{\theta}}[\prod\limits_{t_j:\Delta_{a_i}(t_j)=1}\lambda^A_{\mathcal{O}}(t_j\mid\mathcal{H}^{A}_{it_j},\bm{\theta})\exp\{-\int\lambda^A_{\mathcal{O}}(t\mid\mathcal{H}^{A}_{it},\bm{\theta})dt\}\mid \bm{v}_{1:n} ]}.$$ with a full derivation in \textbf{Supplementary Material B}. The denominator averages over posterior uncertainty in $\bm{\theta}$.

Consequently, one can obtain the optimal interventional parameters through off-the-shelf optimization methods. In this paper, we employ the global optimization algorithm, differential evolution, implemented through the R package \texttt{DEoptim} \citep{JSSv040i06}.

Prior specifications for $\bm{\theta}$ and posterior inferences from $p_{\mathcal{O}}(\bm{\theta}\mid \bm{v}_{1:n})$ proceed in the usual way, the evaluation of weights $w_i(\bm{\eta})$ using Monte Carlo integration requires only a single Markov Chain Monte Carlo (MCMC) sample from these posteriors. The No-U-Turn sampler (NUTS) implemented in the R package \texttt{rstan} \citep{carpenter2017stan,RStan2023} was used for posterior sampling.

\section{Optimal Oxytocin Administration Process for Preventing PPH}
\label{Application}
This section analyzes a real-world oxytocin dataset to estimate the optimal real-time DTR (that is, an oxytocin administration regime) with the goal of reducing postpartum hemorrhage (PPH).

\subsection{Data Source}
\label{Study_Background}

We used data from the Consortium on Safe Labor (CSL) \citep{zhang2010contemporary}, which contains electronic medical record data on maternal characteristics, medical history, prenatal care, labor and delivery, and postpartum outcomes. Our analysis includes $n=1000$ parturients with singleton gestation, vertex presentation, term birth (37-42 weeks), no prior uterine scar, and spontaneous labor onset. These parturients were drawn from five hospitals that used the same labor management protocol and a common data structure. In these hospitals, oxytocin was administered through a pump infusion system.

Outcome, real-time treatment process and covariates included in the analysis are detailed as follows (variable selection follows \cite{zhu2024oxytocin}):

{\it{Outcome}}. The outcome is estimated blood loss (EBLoss; mL), a continuous measure recorded after delivery. We analyzed $Y=\log(\text{EBLoss})$ to improve approximate normality and interpreted results on the original scale via $\exp(Y)$.

{\it{Real-time treatment process}}. The real-time dynamic treatment $A(t)$ is the oxytocin dose stratum at time $t$ ($1$ for upper stratum, $0$ for lower stratum). To evaluate the sensitivity of the proposed method to the threshold, we estimated the optimal interventional parameters under five candidate thresholds: $\{0, 2, 4, 6, 8\}$. These values provide multiple clinically plausible options while maintaining adequate sample size in each stratum for stable estimation.

{\it{Covariates}}. Baseline covariates include nominal variables $\bm{Z}_1$ (maternal demographics, medical and reproductive history, prenatal behaviors, admission status, and labor summary) and continuous/count variables $\bm{Z}_2$ (e.g., maternal age, BMI, cervical dilation at admission, blood pressure, Bishop score). The time-varying covariate $Z_3(t)$ represents cervical dilation measured during labor. \textbf{Supplementary Material C} reports the summary information of baseline covariates and outcome, which shows that some continuous covariates differ significantly in scale. To ensure comparability of variables during model fitting, standardization (scaling) was applied to continuous/count baseline covariates. 

\subsection{Analysis}
\label{Application_Analysis}
Our primary objective is to estimate the optimal real-time oxytocin dosing regimen. This requires developing a clinically interpretable treatment switching intensity model, which necessitates careful selection of covariates. Recent research has demonstrated a significant relationship between maternal BMI and oxytocin dose requirements during labor augmentation. For example, \citet{carlson2017oxytocin} found that maternal BMI accounted for 16.56\% of the variance in hourly oxytocin doses among obese women who delivered vaginally. The study also identified cervical dilation at oxytocin initiation as a significant predictor. While maternal age, gestational age, status of amniotic membranes at hospital admission were not found to be significant factors. Motivated by these findings, the following treatment switching intensity model was considered:
$$\lambda^A_{\mathcal{E}}(t\mid\cdot)=\exp\{\eta_{1}+\eta_{2}Z_{BMI}+\eta_{3}Z_3(t)/10+\eta_{4}(t-T)/20\},$$ where $Z_{\text{BMI}}$ is BMI at admission, $Z_3(t)$ is cervical dilation, and $T$ is the most recent time of a dose-stratum change. The scaling constants $10$ (for cervical dilation) and $20$ (for time since last change) are used to keep covariates on similar numerical scales.

In addition to our proposed method (\textbf{Proposed}), two alternative approaches were considered: (1) Unoptimized method (\textbf{Unopt.}): we calculated the mean EBLoss estimated directly from the observed data, with no policy learning (no optimization); (2) Bayesian Parametric Method (\textbf{BPM}): a Bayesian parametric approach that learns the optimal real-time DTR by generating posterior samples of non-interventional parameters $\bm{\phi}$ and minimizing the expected loss $\int \mathbb{E}_{\bm{V}\sim p(\bm{v}\mid\bm{\eta,\phi})}[\exp(Y)]p(\bm{\phi}\mid \bm{v}_{1:n})d\bm{\phi}$ following \citet{Hua_2022} (details in \textbf{Supplementary Material D}).

It is worth noting that, unlike BPM, the proposed method does not require performing posterior inference on any parameters other than $\bm{\theta}$ in $\lambda^A_{\mathcal{O}}(t\mid \mathcal{H}^A_t,\bm{\theta})$.

\subsection{Results}
\label{Main_Results}
The CSL oxytocin data were analyzed using Unopt., BPM, and Proposed. A total of $4,000$ MCMC iterations were run, with an initial burn-in of $2,000$ iterations and a thinning factor of $4$. Trace plots indicated convergence, as illustrated for $\bm{\theta}$ ($\delta=4$) in \textbf{Supplementary Material E}.

Table \ref{OxyAnalysis} presents the results. Across all candidate thresholds $\delta\in\{0,2,\ldots,8\}$, the proposed method yields lower estimated mean EBLoss under the learned policy than both Unopt.\ and BPM.

\begin{table}[htbp]
	\caption{Oxytocin data: Estimates of optimal interventional parameters and associated mean EBLoss (Est.EBLoss) under different thresholds. For the unoptimized method (Unopt.), \(\hat{\theta}_k\) denotes the posterior mean.}
	\centering
	\begin{tabular}{ccccccc}
		\hline
		\textbf{Method}              & \textbf{Estimates} & \textbf{$\delta=0$} & \textbf{$\delta=2$} & \textbf{$\delta=4$} & \textbf{$\delta=6$} & \textbf{$\delta=8$}\\ \hline
		\multirow{5}{*}{Unopt.} & $\hat{\theta}_1$          & -2.323                    & -2.364                    & -2.622                    & -2.962                    & -3.225\\
		& $\hat{\theta}_2$          & 0.195                     & 0.182                     & 0.390                      & 0.459                     & 0.693\\
		& $\hat{\theta}_3$          & -0.507                    & -0.455                    & -0.466                    & -0.492                    & -0.466\\
		& $\hat{\theta}_4$          & 0.257                     & 0.362                     & 0.382                     & 0.847                     & 0.475\\
		& Est.EBLoss            & 321.079                   & 321.079                   & 321.079                   & 321.079                   & 321.079\\ \hline
		\multirow{5}{*}{BPM}         & $\hat{\eta}^{opt}_{1}$   & -0.724                    & -0.765                    & -1.032                    & -3.072                    & -3.067\\
		& $\hat{\eta}^{opt}_{2}$   & 0.601                     & 0.783                    & -0.716                      & -1.289                    & 0.076\\
		& $\hat{\eta}^{opt}_{3}$   & 1.079                     & 1.060                     & 1.132                     & 0.932                    & -2.062\\
		& $\hat{\eta}^{opt}_{4}$   & 1.847                  & 1.847                     & 1.515                    & -0.824                    & 1.490\\
		& Est.EBLoss            & 267.321                  & 253.640                   & 275.446                   & 285.021                   & 282.010\\ \hline
		\multirow{5}{*}{Proposed}    & $\hat{\eta}^{opt}_{1}$   & -1.523                    & -1.564                     & -1.822                    & -2.013                  & -2.226\\
		& $\hat{\eta}^{opt}_{2}$   & \textbf{0.995}                     & \textbf{0.982}                     & \textbf{1.189}                     & \textbf{1.407}                     & \textbf{1.692} \\
		& $\hat{\eta}^{opt}_{3}$   & 0.293                     & 0.345                     & 0.333                     & 0.456                   & 
		0.528 \\
		& $\hat{\eta}^{opt}_{4}$   & 1.057                    & 1.161                     & 1.181                    & 1.797                     & 1.445\\
		& Est.EBLoss            & \textbf{130.162}                   & \textbf{141.305}                   & \textbf{150.435}                   & \textbf{135.473}                   & \textbf{114.687}                  \\ \hline
	\end{tabular}
	\label{OxyAnalysis}
\end{table}

In addition, the proposed method yields $\hat{\eta}^{opt}_{2}>0$ for all thresholds. Under our intensity model, this means that higher BMI is associated with a higher switching intensity. In clinical terms, parturients with higher BMI are expected to require earlier and more frequent dose-stratum adjustments, including earlier escalation to the high-dose stratum when clinically appropriate.

Moreover, unlike the unoptimized approach, the proposed method yields $\hat{\eta}^{opt}_{3}>0$ for all thresholds. This suggests that, as cervical dilation progresses, the switching intensity increases, consistent with the need for closer monitoring and timely dose titration during active labor. We also find $\hat{\eta}^{\text{opt}}_{4}>0$ under both Unopt.\ and the proposed method, indicating higher switching intensity when more time has elapsed since the last dose-stratum change. Notably, the proposed method produces larger values of $\hat{\eta}^{\text{opt}}_{4}$ than Unopt., suggesting that observed practice is less responsive to long intervals without titration than the policy learned by our method.

In contrast, BPM produces unstable estimates for $\hat{\eta}^{opt}_{2}$, with the sign changing across thresholds (e.g., positive at $\delta=2$ but negative at $\delta=4$, and negative at $\delta=6$ but positive at $\delta=8$). This instability makes the implied clinical interpretation sensitive to the choice of $\delta$ and may reflect model misspecification and/or sensitivity to unmeasured factors.

Overall, the directions implied by the proposed method are consistent with prior findings in \citet{carlson2017oxytocin}. Importantly, our estimates go beyond qualitative statements (e.g., "higher BMI should start earlier") by quantifying how patient factors change the real-time switching intensity of the dosing stratum.

\begin{figure}[htbp]
	\centering
	\includegraphics[width=1\linewidth]{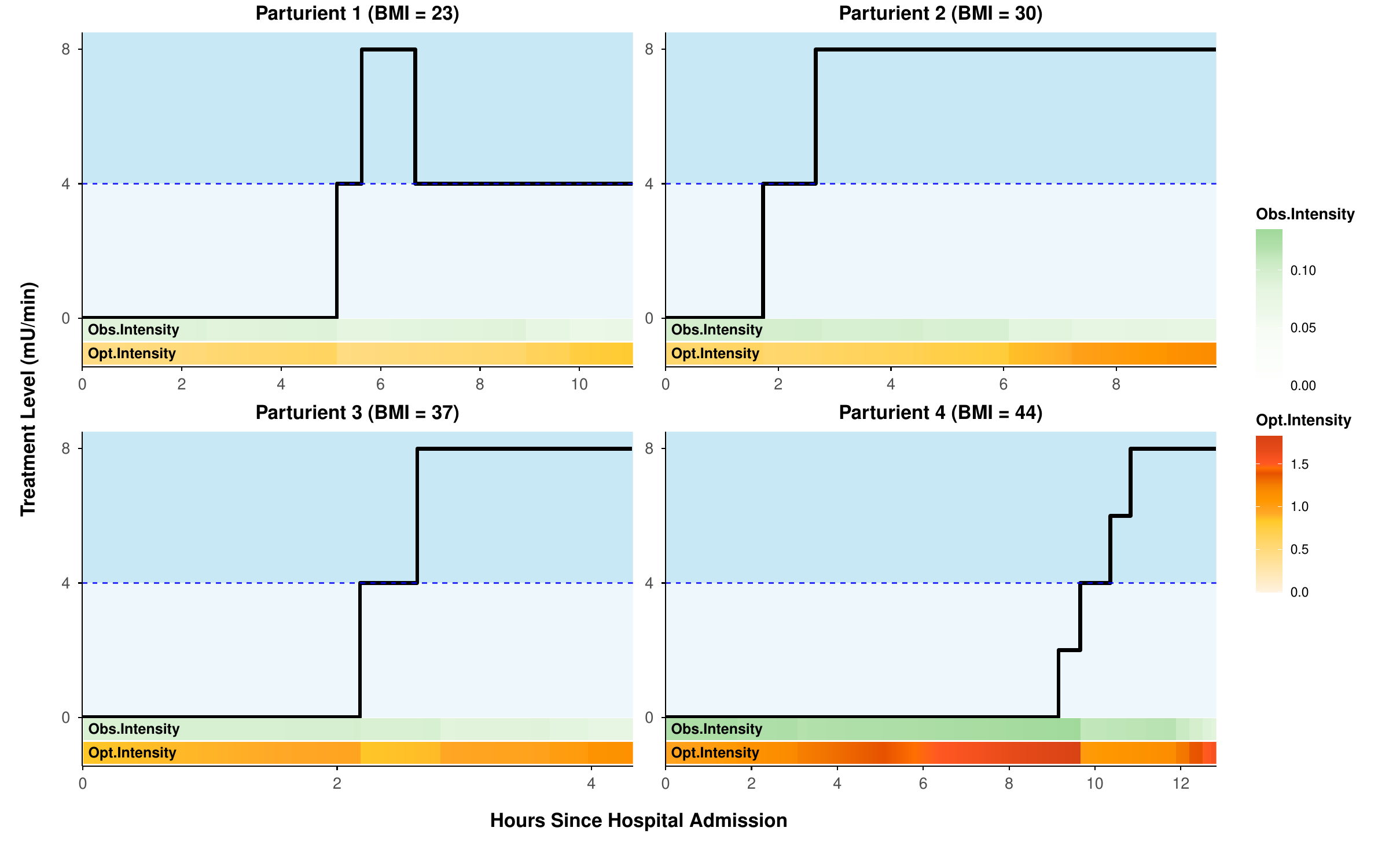}
	\caption{Oxytocin administration trajectories and treatment switching intensities (obs vs. optimized by the proposed method) for four parturients with different BMI levels. Darker colors indicate higher intensity. The blue dashed line marks the threshold $\delta=4$ (median dose range). }
	\label{fig:optimal_treatment_plot}
\end{figure}

To further elucidate the clinical applicability of the proposed approach, Figure~\ref{fig:optimal_treatment_plot} displays observed oxytocin trajectories and the corresponding switching intensities (observed vs. optimized by the proposed method) for four parturients with varying BMI levels. In the absence of prior clinical knowledge, we selected the median of the dose range $[0, 8]$ observed across these four individuals, setting $\delta=4$. In these examples, observed stratum changes are generally less frequent than those suggested by the learned policy. For instance, for parturient 3 (BMI = 37), the learned policy recommends timely de-escalation after reaching peak dose, whereas the observed trajectory remains in the upper dose stratum; for parturient 4 (BMI = 44), the recommended switching intensity peaks over the interval of approximately 8--9.5~h (indicated by the darkest color bar), consistent with the actual switch from the lower to the upper dose stratum. However, while the recommended switching intensity is relatively high during the interval of 4--9~h and after 12~h, no timely dose-stratum adjustment was made in real practice. This underscores the potential for improved individualized real-time dosing strategies.

\section{Simulation}
\label{Simulation}
In this section, simulations are conducted to evaluate the effectiveness of the proposed method in estimating optimal real-time DTRs. Section \ref{Simulation_Scheme} describes the simulation scheme. Section \ref{Data_Generation} details the data generation, and Section \ref{Simulation_Results} presents the simulation results.

\subsection{Simulation Scheme}
\label{Simulation_Scheme}

To closely mimic the real dataset and evaluate the proposed method's performance under unobserved confounding and varying model complexities, simulation studies were performed under four scenarios:

\begin{itemize}[leftmargin=15pt]
	\item Case 1 (No-Un, E-L): No unobserved variables, with the treatment switching intensity model \(\lambda^A_{\mathcal{O}}(t\mid \mathcal{H}^A_t,\bm{\theta})\) as the exponential of a linear combination of variables in \(\mathcal{H}^A_t\);
	\item Case 2 (No-Un, E-NL): No unobserved variables, with \(\lambda^A_{\mathcal{O}}(t\mid \mathcal{H}^A_t,\bm{\theta})\) as the exponential of a nonlinear combination of variables in \(\mathcal{H}^A_t\);
	\item Case 3 (Un, E-L): An unobserved variable $U\in \mathbb{R}$ is present, with $\lambda^A_{\mathcal{O}}(t\mid \mathcal{H}^A_t,\bm{\theta})$ as the exponential of a linear combination of variables in $\mathcal{H}^A_t$;
	\item Case 4 (Un, E-NL): An unobserved variable $U\in \mathbb{R}$ is present, with \(\lambda^A_{\mathcal{O}}(t\mid \mathcal{H}^A_t,\bm{\theta})\) as the exponential of a nonlinear combination of variables in \(\mathcal{H}^A_t\).
\end{itemize}
Here, "No-Un" denotes no unobserved variables and “Un” denotes unobserved variables, while “E”, “L”, and “NL” represent exponential, linear, and nonlinear, respectively. 

Due to the limited availability of comparable methods, we primarily considered Unopt., BPM, and the proposed method. For each simulation replicate, we evaluated each method using the following metrics:
\begin{itemize}[leftmargin=15pt]
	\item Mean loss: the average loss computed by 2,000 independent individuals following the estimated optimal policy.
	\item Running time: the total time for posterior sampling and the optimization step.
\end{itemize}
We repeated this procedure for 100 independent simulation replicates and reported the mean and standard deviation of each metric. We considered three sample sizes in each setting: $n\in\{200,600,1000\}$.

\subsection{Data Generation}
\label{Data_Generation}
The data-generating processes are described below, with $t\in [0,3]$ for all settings:

\textbf{Case 1 (No-Un, E-L):} In this scenario, baseline covariates consist of binary variables $\bm{Z}_1=(Z_{1,1},\ldots,Z_{1,30})^\top$ and continuous variables $\bm{Z}_2=(Z_{2,1},\ldots,Z_{2,10})^\top$:
\begin{gather*}
	Z_{1,k}\sim \text{Bernoulli}(p_k), \text{ where } p_k\sim \text{Uniform}(0.1,0.9),k=1,...,30\\
	Z_{2,k}\sim \mathcal{N}(\mu_k,\sigma^2_k),\text{ where }\mu_k\sim \mathcal{N}(0,0.5^2),\sigma_k\sim \text{Uniform}(0.1,1),k=1,...,10
\end{gather*}
Let $\tilde{Z}_1=\frac{1}{30}\sum_{k=1}^{30}Z_{1,k}$ and $\tilde{Z}_2=\frac{1}{10}\sum_{k=1}^{10}Z_{2,k}$ denote the corresponding averages. The time-varying covariate $Z_3(t)$ is observed on the grid $t\in\{0,0.1,\ldots,3\}$ and evolves as
\[
Z_3(t\mid\cdot)\sim \mathcal{N}(1+0.1\int_0^{t^-} A(s)ds+0.025Z_3(t^-)+0.01\tilde{Z}_1+0.01\tilde{Z}_2,1)\text{ }\text{for}\text{ }t\in \{0,0.1,\dots,3\}.
\]
Treatment completion is generated from a counting process with intensity
\[
	\lambda^{T_{max}}(t\mid \cdot)=\exp\{0.1+0.05A(t^-)+0.06\tilde{Z}_1-0.1\tilde{Z}_2-0.05Z_3(t^-)\}.
\]
Treatment stratum changes follow the observational switching intensity
\[
\lambda^A_{\mathcal{O}}(t\mid \cdot)=\exp\{-0.1+0.05Z_3(t)+0.1(t-T)\},
\]
where $T$ is the most recent time of a stratum change prior to $t$. The outcome is generated as
\[
Y\sim \mathcal{N}(3-0.1J+0.1\int Z_3(t)dt+\bm{\phi}^\top_{Y1}\bm{Z}_{1}+\bm{\phi}^\top_{Y2}\bm{Z}_{2},0.5^2),
\]
where $J$ is the total number of stratum changes over $[0,T_{\max}]$ and
$(\bm{\phi}^\top_{Y1},\bm{\phi}^\top_{Y2})^\top\sim \mathcal{N}(\bm{0},0.1^2\bm{I}_{40\times 40})$. $\bm{I}_{40\times 40}$ is the $40\times 40$ identity matrix. The full covariate vector at time $t$ is $\bm{L}(t)=(\bm{Z}_1^\top,\bm{Z}_2^\top,Z_3(t))^\top$.

\textbf{Case 2 (No-Un, E-NL):} In this scenario, $\bm{L(t)}$, $\lambda^{T_{max}}$, and $(\bm{\phi}_{Y1}^\top,\bm{\phi}_{Y2}^\top)^\top$ are generated in the same manner as in Case 1 (No-Un, E-L), but the treatment switching intensity model and outcome model are nonlinear as follows:
\begin{gather*}
	\lambda^A_{\mathcal{O}}(t\mid \cdot)=\exp\{ [0.8A(t^-)-0.5(1-A(t^-))][2S(Z_3(t);1.2)-1] \}\\
	Y\sim \mathcal{N}(1-4\int A(t)\{2S(Z_3(t);1.5)-1\}dt+\bm{\phi}_{Y1}^\top\bm{Z}_{1}+\bm{\phi}_{Y2}^\top\bm{Z}_{2},0.1^2), 
\end{gather*}
where $S(Z_3(t);k)=\frac{1}{1+\exp(-10(Z_3(t)-k))}$ denotes a scaled and shifted sigmoid function. The threshold parameter $k$ is set to $1.2$ in \(\lambda^A_{\mathcal{O}}(t\mid \cdot)\) and to $1.5$ in the outcome model for $Y$.

In Case 2 (No-Un, E-NL), \(\lambda^A_{\mathcal{O}}(t\mid \cdot)\) is specified to account for both the binary treatment state effect and the nonlinear impact of \(Z_3(t)\)  (via transformed sigmoid), mimicking a complex scenario.


\textbf{Case 3 (Un, E-L)} In this scenario, we introduce an unobserved variable $U\sim \mathcal{N}(1,1)$ that affects intermediate covariates and the outcome. Data are generated as follows:
\begin{gather*}
	Z_{1,k}\sim \text{Bernoulli}(p_k), \text{ where } p_k\sim \text{Uniform}(0.1,0.9),k=1,...,30\\
	Z_{2,k}\sim \mathcal{N}(\alpha_k+\beta_kU,\sigma_k^2),\text{ where }\alpha_k\sim \mathcal{N}(0,0.5^2),\beta_k\sim \mathcal{N}(1,0.5^2),\sigma_k\sim \text{Uniform}(0.1,1),k=1,...,10\\
	Z_3(t\mid\cdot)\sim \mathcal{N}(1+0.1\int_0^{t^-} A(s)ds+0.025Z_3(t^-)+0.01\tilde{Z}_1+0.01\tilde{Z}_2+0.01U ,1)\text{ }\text{for}\text{ }t\in \{0,0.1,\dots,3\}\\
	\lambda^{T_{max}}(t\mid \cdot)=\exp\{0.1+0.05A(t^-)+0.06\tilde{Z}_1+0.08\tilde{Z}_2-0.05Z_3(t^-)+0.07U\}\\
	\lambda^A_{\mathcal{O}}(t\mid \cdot)=\exp\{-0.1+0.05Z_3(t)+0.1(t-T)\}\\
	Y\sim \mathcal{N}(3-0.1J\tilde{Z}_2+0.1\int Z_3(t)dt+\bm{\phi}_{Y1}^\top \bm{Z}_{1}+0.6U,0.3^2),\text{ where }\bm{\phi}_{Y1}\sim \mathcal{N}(\bm{0},0.1^2\bm{I_{30\times 30}}),
\end{gather*}
where $\bm{I}_{30\times 30}$ is the $30\times 30$ identity matrix and other notation (e.g., $\tilde{Z}_1$, $\tilde{Z}_2$, $T$, and $J$) follows Case 1.

\textbf{Case 4 (Un, E-NL)} In this scenario, $U$, $\bm{L(t)}$, $\lambda^{T_{max}}$, and $\bm{\phi}_{Y1}$ are generated in the same manner as in Case 3 (Un, E-L), but treatment switching intensity model and outcome model are nonlinear:
\begin{gather*}
	\lambda^A_{\mathcal{O}}(t\mid \cdot)=\exp\{ [0.8A(t^-)-0.5(1-A(t^-))](2S(Z_3(t);1.2)-1) \}\\
	Y\sim \mathcal{N}(1-3\tilde{Z}_2\int A(t)\{2S(Z_3(t);1.5)-1\}dt+\bm{\phi}_{Y1}^\top \bm{Z}_{1}+0.6U,0.3^2) 
\end{gather*}
Note that in Case 3 (Un, E-L) and Case 4 (Un, E-NL), $U$ affects $\bm{Z}_2$. Since the treatment effect component of $Y$ depends on $\tilde{Z}_2$ (through $J\,\tilde{Z}_2$ in Case~3 and through $\tilde{Z}_2\int A(t)\{2S(\cdot)-1\}dt$ in Case~4), $U$ induces confounding through its impact on $\bm{Z}_2$ and $Y$. As a result, methods that omit $U$ (such as BPM as implemented here) can be sensitive to this latent structure, leading to biased parameter learning and potentially suboptimal estimates of the interventional parameters.

To generate the aforementioned variables, especially the real-time treatment process, a standard discrete-time approximation is used (cf. \citet{penichoux2015simulating}). We partition $[0,t_R]$ into intervals of a small length $dt$ (viewed as a precision parameter; set to 0.01 in both simulations and real data analyses), which controls the accuracy of integration and the discreteness of the time interval. The occurrence of changing the treatment schedule during the interval $[t,t+dt)$ is generated with an event probability $\lambda^A_{\mathcal{O}}(t\mid\cdot)dt$. This approximation allows us to update treatment and covariate histories sequentially and to compute integrals (e.g., $\int_0^{T_{\max}}A(s)\,ds$) by numerical summation on the same grid. Full algorithmic details are provided in \textbf{Supplementary Material F}.

\subsection{Simulation Results}
\label{Simulation_Results}
For each case, we estimated an optimal policy by optimizing the parameters in the experimental-world switching intensity $\lambda^A_{\mathcal{E}}(t\mid\cdot,\bm{\eta})$. We used the following policy classes: 
\begin{itemize}[leftmargin=15pt]
	\item Case 1 (No-Un, E-L) and Case 3 (Un, E-L): $\lambda^A_{\mathcal{E}}(t\mid \cdot,\bm{\eta})=\exp\{\eta_1+\eta_2Z_3(t)+\eta_3(t-T)\}$;
	\item Case 2 (No-Un, E-NL) and Case 4 (Un, E-NL): $\lambda^A_{\mathcal{E}}(t\mid \cdot,\bm{\eta})=\exp\{ [\eta_1A(t^-)+\eta_2(1-A(t^-))][2S(Z_3(t);\eta_3)-1]\}$.
\end{itemize}

Table \ref{Case12_simres} summarizes the results of Case 1 (No-Un, E-L) and Case 2 (No-Un, E-NL), reporting the average and standard deviation of the estimated mean losses across 100 replications, as well as the mean and standard deviation of the running time (RT) in seconds for the MCMC and optimization procedure.

As shown in Table \ref{Case12_simres}, BPM and the proposed method yield significant reductions in loss compared to the unoptimized method under both the Case 1 (No-Un, E-L) and Case 2 (No-Un, E-NL). The average loss reductions are similar for BPM and the proposed method, while the proposed method has the smallest loss standard deviation in Case 1, and shows a slightly higher SD than BPM in Case 2. As the sample size increases, the proposed method shows a decreasing mean loss and a modest decrease in variability.

A main practical benefit of the proposed method is the computational efficiency. It is consistently faster than BPM across sample sizes and settings. For example, in Case 1 with $n=600$, the average runtime decreases from 978.79 seconds for BPM to 401.66 seconds for the proposed method (about a 59\% reduction). The proposed method also shows smaller runtime variability than BPM, which makes computation more predictable. Because Unopt.\ does not involve posterior sampling or optimization, we do not report runtimes for that method (shown as NA).

\begin{table}[htbp]
	\caption{Case 1 (No-Un, E-L) and Case 2 (No-Un, E-NL) results: estimated mean loss (Loss) under the estimated optimal policy (average with standard deviation) and total runtime in seconds (RT, average with standard deviation) of MCMC and optimization across varying sample sizes (n) and settings, based on 100 simulation runs.}
	\begin{tabular}{llllll}
		\hline
		&        & \multicolumn{2}{c}{Case 1 (No-Un, E-L)} & \multicolumn{2}{c}{Case 2 (No-Un, E-NL)} \\ \hline
		n & Method & Loss (SD)      & RT (SD)      & Loss (SD)      & RT (SD)      \\ \hline
		\multirow{3}{*}{200} & Unopt. & 31.95(1.25) & NA              & 14.18(6.11) & NA              \\
		& BPM        & 25.12(1.88) & 758.22(66.71)  & 5.52(2.74) & 522.88(244.09) \\
		& \textbf{Proposed}    & \textbf{22.35(0.43)} & \textbf{177.80(28.83)}    & \textbf{6.03(2.87)} & \textbf{54.94(6.38)}    \\ \hline
		\multirow{3}{*}{600} & Unopt. & 31.99(0.73) & NA              & 14.07(3.38) & NA             \\
		& BPM        &24.56(1.37) & 978.79(215.60) & 4.81(1.58) & 2367.31(1549.14)  \\
		& \textbf{Proposed}    & \textbf{22.35(0.41)} & \textbf{401.66(78.75)}   &\textbf{5.49(2.61)} & \textbf{158.14(19.79)}   \\ \hline
		\multirow{3}{*}{1000} & Unopt. & 32.01(0.63) & NA             & 14.44(2.80) & NA              \\
		& BPM         & 24.44(1.14) & 1309.59(210.62) & 5.04(1.87)& 2741.87(1136.87) \\
		& \textbf{Proposed}    &\textbf{22.34(0.41)} & \textbf{755.97(135.78)}  & \textbf{5.15(2.31)} & \textbf{239.60(36.83)}  \\ \hline
	\end{tabular}
	\label{Case12_simres}
\end{table}
\begin{table}[htbp]
	\caption{Case 3 (Un, E-L) and Case 4 (Un, E-NL) results: estimated mean loss (Loss) under the estimated optimal policy (average with standard deviation) and total runtime in seconds (RT, average with standard deviation) of MCMC and optimization across varying sample sizes (n) and settings, based on 100 simulation runs.}
	\begin{tabular}{cccccc}
		\hline
		&        & \multicolumn{2}{c}{Case 3 (Un, E-L)} & \multicolumn{2}{c}{Case 4 (Un, E-NL)} \\ \hline
		n & Method & Loss (SD)      & RT (SD)      & Loss (SD)      & RT (SD)      \\ \hline
		\multirow{3}{*}{200} & Unopt. & 79.39(4.31) & NA             & 112.77(265.53) & NA             \\
		& BPM         & 82.76(2.84) & 728.08(81.14)  & 2294.76(3305.41) &560.59(204.03) \\
		& \textbf{Proposed}    & \textbf{57.65(1.74)} & \textbf{89.41(11.76)}    & \textbf{60.02(64.53)} & \textbf{121.11(20.94)}   \\ \hline
		\multirow{3}{*}{600} & Unopt. & 79.45(2.15) & NA             & 375.29(1439.42) & NA              \\
		& BPM        & 82.95(2.22) &938.85(83.32)  & 2875.19(3375.45) & 1706.99(576.41) \\
		& \textbf{Proposed}    & \textbf{57.42(1.17)} & \textbf{239.17(22.97)}  & \textbf{36.81(48.25)} & \textbf{317.49(39.08)}   \\ \hline
		\multirow{3}{*}{1000} &Unopt.& 79.22(1.75)& NA             & 295.41(923.41) & NA              \\
		& BPM        & 83.31(1.78) & 1405.30(243.92) & 3048.01(3970.95) & 3053.72(1080.74) \\
		& \textbf{Proposed}    & \textbf{57.29(1.07)} & \textbf{450.95(46.07)} & \textbf{30.32(42.42)} & \textbf{393.08(33.82)}  \\ \hline
	\end{tabular}
	\label{Case34_simres}
\end{table}

Table~\ref{Case34_simres} presents the results for Case 3 (Un, E-L) and Case 4 (Un, E-NL). As in Table~\ref{Case12_simres}, we summarized the mean and standard deviation of the estimated mean loss (over 100 replicates) and the mean and standard deviation of total runtime for posterior sampling and optimization. Notably, the proposed method achieves the lowest mean loss and the shortest runtime among the methods considered. As sample size increases, the mean loss under the proposed method decreases and its variability also decreases. In contrast, BPM performs poorly in these two cases and can be worse than Unopt. This behavior is expected in our design because BPM, as implemented here, omits the latent variable $U$ even though $U$ affects intermediate covariates and the outcome. This suggests that BPM can learn an incorrect policy when the parametric models are misspecified with respect to this latent structure.

Overall, the proposed method improves on Unopt.\ in all settings. When $U$ is absent (Cases~1--2), its loss reduction is comparable to BPM, while remaining faster. When $U$ is present (Cases~3--4), the proposed method shows a clearer advantage in loss, and it remains substantially more computationally efficient than BPM across all scenarios.
\section{Discussion}
\label{Discussion}
Motivated by the optimization of the oxytocin administration process, we proposed a semiparametric Bayesian approach for learning optimal real-time DTRs, and developed a "physician-in-the-loop" dose recommendation framework. The aim is to support personalized medication decisions and the scientific management of chronic diseases.

In simulations, the proposed method achieves lower loss and is consistently faster than the competing methods, with the largest advantage when latent factors affect intermediate covariates and outcomes. In the CSL application, the learned switching-intensity parameters provide direct clinical interpretation. Across thresholds, higher baseline BMI is associated with earlier initiation, earlier escalation to higher dose levels, and more frequent titration checks under the learned policy. Increasing cervical dilation is also associated with higher switching intensity, consistent with closer monitoring and timely titration during active labor. In addition, longer elapsed time since the last dose-stratum change is linked to stronger subsequent adjustments. These patterns are stable across the threshold grid and provide a transparent mapping from estimated parameters to actionable guidance.

This study makes modeling choices to balance interpretability and implementability. For the application, the continuous infusion process is represented by discrete dose levels to match current clinical workflows, which improves operational feasibility but may not capture finer-scale decisions that a fully continuous policy could provide. We also use parametric specifications for the switching intensities to enhance interpretability and facilitate deployment; however, a specific parametric form may not generalize to all settings. Finally, the CSL oxytocin records cannot be publicly shared for confidentiality reasons. We provide code and supplementary documentation to support reproduction, but end-to-end replication still requires access to similar data or appropriate approvals.

There are two potential extensions. First, in this study, the treatment level at each time point is binary, meaning that the recommended optimal dose is categorized into two regions (upper stratum vs. lower stratum). Future work could extend this framework by modeling treatment levels as multi-category, enabling finer stratification of optimal dose. For instance, under a three-level classification, the optimal recommendations could be interpreted as low, moderate, and high dose levels. For multi-valued cases, models based on filtered counting processes, as discussed in \citet{Hu_et_al_2023}, can be developed. Second, nonparametric treatment switching intensity models (e.g., neural networks or Gaussian processes) would be a straightforward next step, offering the potential to learn more complex dependence on patient history while retaining the same two-world, importance-weighted decision framework.

\section*{Supplementary Materials}
\textbf{Supplemental materials A,B,C,D,E,F.} (PDF file)\\
\textbf{R codes} for implementing the simulation studies can be found from the GitHub link \href{https://github.com/Haiyan-Codes/BNP-CTMSM}{https://github.com/Haiyan-Codes/BayesianLearning-of-RealTimeDTRs}.\\
Oxytocin data are not shared due to confidentiality.

\section*{Disclosure Statement}\label{disclosure-statement}
The authors report no potential conflict of interest.

\section*{Funding}
This study is funded by the Shanghai Natural Science Foundation (24ZR1420400), National Natural Science Foundation of China (12401347, 12571283 and 72331005), Fundamental Research Funds for the Central Universities, and the Shanghai "Science and Technology Innovation Action Plan" Computational Biology Key Program (23JS1400500 and 23JS1400800).

\bibliography{bibliography.bib}
\end{document}